\documentclass[
aps,prl,reprint,superscriptaddress,amsmath,amssymb,longbibliography
]{revtex4-2}

\usepackage{subcaption}
\usepackage{bbold}
\usepackage{ragged2e} 
\DeclareCaptionJustification{justified}{\justifying}
\DeclareMathOperator*{\argmin}{arg\,min}

\usepackage{graphicx}
\graphicspath{{figures/}}

\usepackage{dcolumn}
\usepackage{bm}
\usepackage{braket}
\usepackage{enumitem}

\usepackage[dvipsnames, table]{xcolor}

\usepackage{orcidlink}
\usepackage{hyperref}
\def\colorLinks{black}
\def\colorUrl{blue}
\def\colorCitations{blue}
\hypersetup{
  colorlinks=true,
  linkcolor=\colorLinks,
  urlcolor=\colorUrl,
  citecolor=\colorCitations
}

\usepackage{hyperxmp}
\usepackage{lipsum}

\usepackage{algorithm2e}
\RestyleAlgo{ruled}
\usepackage{algpseudocode}
\SetKwComment{Comment}{// }{}

\usepackage[normalem]{ulem}

\def\Title{Phase transitions in quantum-circuit compilation}

\begin{document}

\preprint{APS/123-QED}

\title{\Title}

\author{Andrea De Girolamo\orcidlink{0009-0002-4529-0139}}\email{andrea.degirolamo@phd.unipd.it}
\affiliation{Dipartimento di Fisica e Astronomia ``Galileo Galilei'', Università degli Studi di Padova, 35131, Padova, Italy}
\affiliation{Istituto Nazionale di Fisica Nucleare, Sezione di Padova, Padova, Italy}

\author{Davide Rattacaso\orcidlink{0000-0001-8219-5806}}
\affiliation{Dipartimento di Fisica e Astronomia ``Galileo Galilei'', Università degli Studi di Padova, 35131, Padova, Italy}
\affiliation{Istituto Nazionale di Fisica Nucleare, Sezione di Padova, Padova, Italy}

\author{\\Simone Notarnicola\orcidlink{0000-0001-8364-6330}}
\affiliation{Dipartimento di Fisica e Astronomia ``Galileo Galilei'', Università degli Studi di Padova, 35131, Padova, Italy}
\affiliation{Istituto Nazionale di Fisica Nucleare, Sezione di Padova, Padova, Italy}

\author{Ilaria Siloi\orcidlink{0000-0002-3806-2034}}
\affiliation{Dipartimento di Fisica e Astronomia ``Galileo Galilei'', Università degli Studi di Padova, 35131, Padova, Italy}
\affiliation{Istituto Nazionale di Fisica Nucleare, Sezione di Padova, Padova, Italy}
\affiliation{Padua Quantum Technologies Research Center, Università degli Studi di Padova, Italy}

\author{Simone Montangero\orcidlink{0000-0002-8882-2169}}
\affiliation{Dipartimento di Fisica e Astronomia ``Galileo Galilei'', Università degli Studi di Padova, 35131, Padova, Italy}
\affiliation{Istituto Nazionale di Fisica Nucleare, Sezione di Padova, Padova, Italy}
\affiliation{Padua Quantum Technologies Research Center, Università degli Studi di Padova, Italy}

\date{\today}

\begin{abstract}

Quantum-circuit compilation aims at finding an optimized realization of a target circuit under given constraints, e.g., the minimization of hardware-induced errors and the unitary-equivalence of the circuit. We connect the compilation process with the thermodynamics of a many-body spin system: circuit infidelity plays the role of the energy function and low-temperature states correspond to compiled circuits. In the paradigmatic case where crosstalk between parallel gates is present, we find a phase transition between a disordered phase and an antiferromagnetic brick-wall phase, compatible with the Ising universality class. At larger crosstalk, we observe a \mbox{$\mathbb{Z}_3$-ordered} regime, suggesting that increasingly serial compiled circuits are associated with emergent $\mathbb{Z}_n$-ordered phases. When the unitary-equivalence constraint is removed, these phases disappear, showing that the equivalence between circuits underlies the emergent criticality and constitutes a source of complexity in quantum-circuit compilation and, more generally, in equivalence-constrained optimization. Finally, we observe that the Kolmogorov complexity of the circuit enhances the emergence of ordered phases.

\end{abstract}

\maketitle

Optimization problems can undergo sharp changes in the organization of their low-energy configurations. As a control parameter is varied, near-optimal states may reorganize into distinct sectors, separated by large barriers under the local moves used to explore configuration space. In random constraint-satisfaction problems, satisfiability is the canonical example: changing the constraint density can reorganize the solution space into clustered regions, thereby producing sharp changes in typical algorithmic hardness~\cite{Cook1971NP, Nemhauser1988CombOp, Karp1972CombOptNPComplete, Cheeseman1991HardProblems, Mitchell1992DistributionsSAT, Kirkpatrick1994CriticalRandomBoolean, Monasson1999ComplexityPT, Anderson1999ProblemsFiniteTime, Gomes2002SatisfiedwithPhysics, Mezard2002RandomSatisfiability}. Statistical mechanics provides a natural language for this phenomenon~\cite{Onsager1944Crystal, Kadanoff1967CritPoints, Wilson1974RG, Nishimori2010PTCriticalPhenomena, Kadanoff1966ScalingIsing}. Each admissible configuration is interpreted as a many-body state, and the objective function, or cost, as its energy~\cite{Vannimenus1984statmechtravsalesman, Fu1986StatMechNPOptimization, SherringtonKirkpatrickModelSpinGlass, Nishimori2001StatphysSpinGlass, Martin2001StatMechOptimizationProblems, Lucas2014IsingNP}. An artificial temperature then defines a Boltzmann ensemble over configurations, progressively biased toward low-cost states as the temperature is lowered~\cite{Kirkpatrick1983SimAnnealing}. This ensemble probes not only the optimum, but also the organization of near-optimal configurations, distinguishing disordered, clustered, and ordered regimes. This viewpoint underlies simulated annealing (SA) and its quantum extensions, including adiabatic quantum computation and quantum annealing~\cite{Kadowaki1998QAnnealingTFIM, Farhi2000QCAdiabaticEvolution, Farhi2001QAdiabaticNPComplete, Santoro2002QAnnealingIsingSpinGlass, Albash2018AdiabaticQC, Hauke2020QAnnealingPerspectives}.

Here we show that an equivalence constraint can organize the low-energy sector of an optimization landscape~\cite{Huet1980EquationsRewriteRules, Plaisted1993EqReasoning, Rattacaso2026QAlgos4EqReas, Clement2023EqReasQC, Clement2024EqReasQCMinimal, Blake2026SimplerEqReasQC, Rattacaso2026EqReasPolymers}. Quantum-circuit compilation provides a minimal setting~\cite{Chong2017Compiler4QHardware, Wille2009EqCheckCircuits, Burgholzer2021AdvEqCheckQC, Peham2022EqCheckQCZXCalc, Sander2025EqCheckQCMPO, Schmid2024NAQC, Sivarajah2021TKet, Zhou2020QCTransfSimAnneal, Maronese2022QCompiling, Wetering2025optimalcompilation}: among all circuits that implement a fixed unitary transformation, one seeks a representative that minimizes a hardware-dependent cost, here the expected infidelity on a target noisy device~\cite{Schmid2024NAQC, Rattacaso2025QQCompiling}. The constraint is essential. If the same cost were minimized over all circuits, the optimum would be the empty circuit. Instead, if restricted to a fixed unitary equivalence class, it defines a nontrivial statistical ensemble whose low-temperature states can develop long-range order.

\begin{figure}
    \centering
    \includegraphics[width=\linewidth]{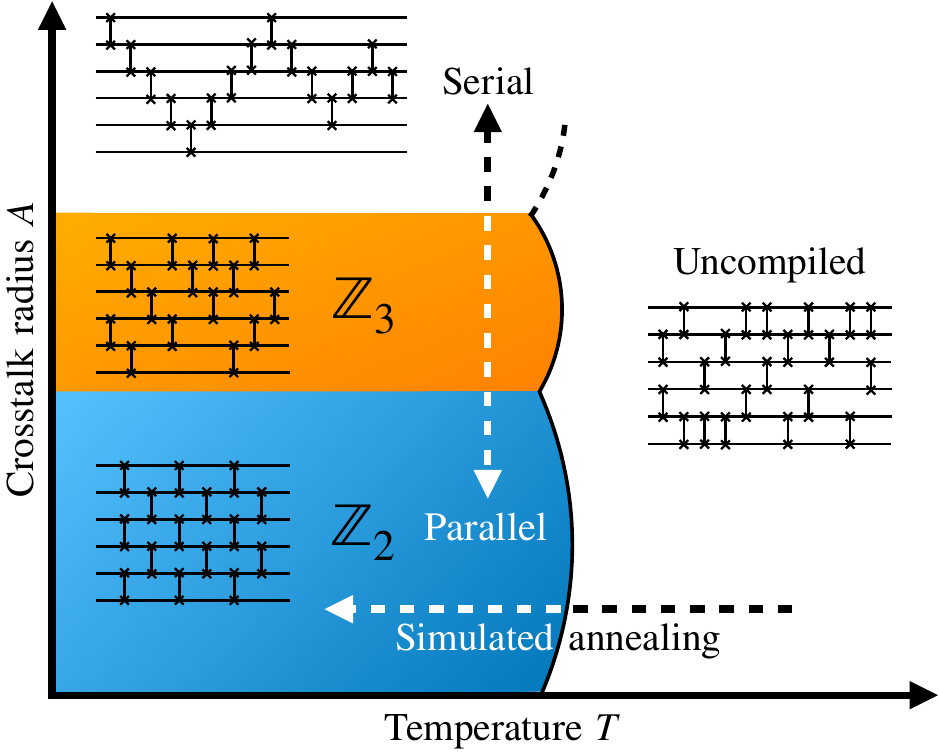}
    \caption{Phase diagram emerging from quantum-circuit compilation. At high temperatures, a disordered phase is observed in which all uncompiled realizations of the same circuit occur with equal probability. As the temperature is reduced through SA, a phase transition toward compiled circuits occurs. As the crosstalk radius increases, compiled circuits display successive $\mathbb{Z}_n$-ordered phases corresponding to increasing serialization.}
    \label{fig:figure1}
\end{figure}
We implement this constrained optimization using the framework of Ref.~\cite{Rattacaso2025QQCompiling}. Starting from a target circuit, local rewrite rules generate alternative circuits that implement the same unitary while changing the gate sequence and circuit depth. We use SA with unitary-preserving rewrite moves to explore the fixed-unitary equivalence class, using a hardware-dependent cost that penalizes circuit depth and same-layer crosstalk. 
Lowering the temperature shifts the underlying Boltzmann distribution, narrowing the canonical ensemble from a flat distribution over all the equivalent circuits toward an ensemble of short, low-crosstalk circuits.

We study the resulting phase diagram as a function of temperature and crosstalk radius (see Fig.~\ref{fig:figure1}). For the circuit that inverts the qubit ordering, compiled into nearest-neighbor $\mathtt{SWAP}$ gates, weak crosstalk produces a transition from a disordered ensemble of equivalent circuits to an antiferromagnetically ordered brick-wall phase. This ordered phase corresponds to a maximally parallel arrangement of $\mathtt{SWAP}$ gates compatible with the target permutation. From finite-size scaling analysis, we find that this transition falls into the two-dimensional (2D) Ising universality class. At larger crosstalk radius, the low-temperature circuits become progressively more serial and show signatures of higher-period order, including a $\mathbb{Z}_3$-ordered regime. We then generalize this analysis by adding single-qubit gates and by considering random circuits built from $\mathtt{SWAP}$ and $\mathtt{CZ}$ gates.

\paragraph{Quantum-circuit compilation.}Given a gate set $\mathbb{G}$ and a register of $N_q$ qubits, we define a circuit $\mathbf{C}$ as a set of instructions $g=\mathtt{G}_{t, q}$, indicating that the gate $\mathtt{G} \in \mathbb{G}$ is applied to the qubit(s) $q$ at the time layer $t$. The corresponding unitary is $\displaystyle \mathcal{U}(\mathbf{C}) = \prod_{t = 1}^{N_t} \bigotimes_{\substack{\mathtt{G}_{t', q} \in \mathbf{C}}}\mathtt{G}\rvert_{t'=t}$, where the product over the $N_t$ layers is time-ordered. We define a parallel execution as the simultaneous execution of instructions acting on disjoint sets of qubits, within the same layer $t$.

We define the infidelity of circuits via the cost function~\cite{Rattacaso2025QQCompiling}
\begin{equation}
    \label{eq:infidelity}
    I(\mathbf{C})= \sum_{\mathtt{Idle}_{t,q} \in \mathbf{C}}i_{\mathtt{Idle}} +\sum_{\mathtt{G}_{t, q} \in \mathbf{C} }i_{\mathtt{G}} + \sum_{\substack{
    \mathtt{G}_{t, q} \in \mathbf{C}\\
    \mathtt{G'}_{t,q'} \in \mathbf{C}\\ q' \neq q}}x_{\mathtt{G}\mathtt{G}'}(\lVert q-q'\rVert) , 
\end{equation}
where $i_{\mathtt{G}}$ is the infidelity associated to the execution of the instruction \mbox{$\mathtt{G}_{t, q}$}, and $i_{\mathtt{Idle}}$ is the error assigned to each inactive qubit $q$ at layer $t$. Additionally, \mbox{$x_{\mathtt{G}\mathtt{G}'}(\lVert q-q'\rVert)$} models the crosstalk error between simultaneous two-qubit gates. Inspired by neutral-atom platforms, we set \mbox{$x_{\mathtt{G}\mathtt{G}'}\left(A, \lVert q - q' \rVert\right) = \left(\dfrac{A}{\lVert q - q' \rVert}\right)^6$}, while crosstalk errors arising from single-qubit gates can be neglected~\cite{Bluvstein2024ReconfAtomArray, Schmid2024NAQC, Rattacaso2025QQCompiling}. This penalizes pairs of two-qubit gates separated by less than the crosstalk radius $A$, while allowing close-packing arrangements at small $A$. First, we notice that $x_{\mathtt{G}\mathtt{G}'}\left(A, \lVert q - q' \rVert\right)$ is independent on the entangling gates $\mathtt{G}$ and $\mathtt{G'}$, allowing us to investigate different classes of circuits. Second, this choice does not limit the generality of this approach, which can be adapted to different platforms, e.g., superconducting qubits and trapped ions~\cite{Zhao2022CrosstalkAnalysisSuperconducting, Wesdorp2026IQMCrosstalkSuperconducting, Nigg2014QCTopoEncodedQubit, ParradoRodriguez2021CrosstalkQECTrappedIons, Fang2022Crosstalk2QGTrappedIons, Gicev2026CrosstalkReview}, with a proper crosstalk error term or other relevant terms.

We fix $i_{\mathtt{G}}$ and $i_{\mathtt{Idle}}$ with $i_{\mathtt{G}} > i_{\mathtt{Idle}}$, keeping the crosstalk radius $A$ as a free parameter. Henceforth, we will indicate the infidelity as $I(A,\mathbf{C})$. Given a circuit $\mathbf{C}$ implementing the unitary $\mathcal{U}(\mathbf{C})$, we call $\mathcal{C}_{\mathcal{U}(\mathbf{C})}$ its equivalence class, namely the set containing all the equivalent circuits, i.e., the circuits that implement the same unitary. If $I(A,\mathbf{C})$ represents the infidelity error, \textit{compilation} refers to the process of identifying $\mathbf{C_*} = \argmin_{\mathbf{C'} \in \mathcal{C_{\mathcal{U}(\mathbf{C})}}} I(A^*,\mathbf{C'})$, namely the equivalent circuit that minimizes the infidelity.

\begin{figure}[t]
    \centering
    \includegraphics[width=\linewidth]{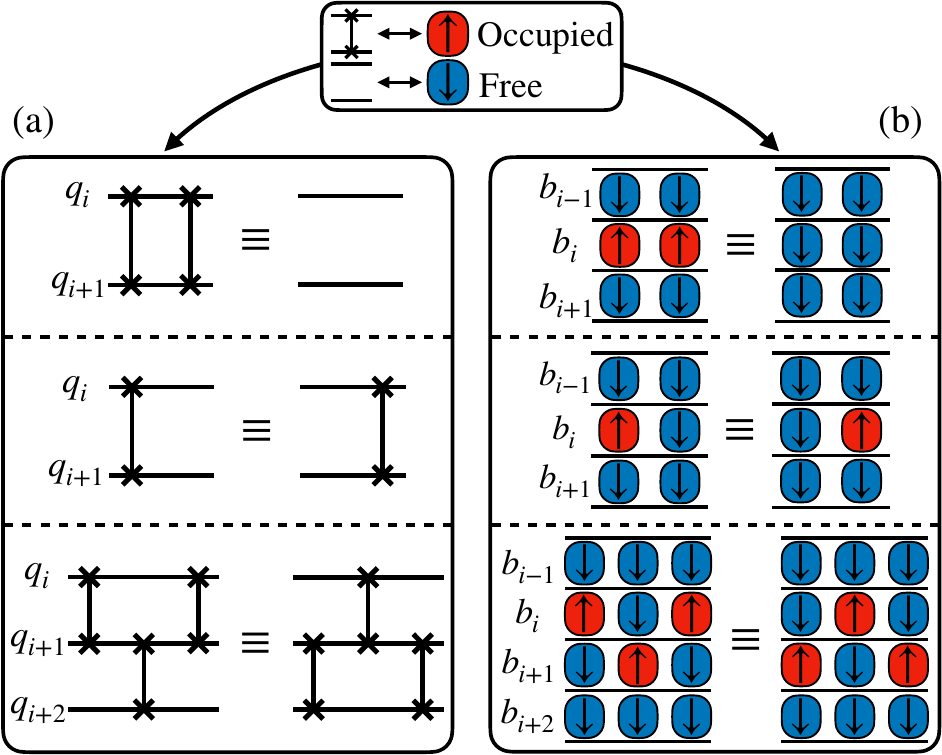}
    \caption{(a)~Equivalence rules for circuits with $\mathtt{SWAP}$ gates. From top to bottom, the rules represent mutual cancellation, time shift, and rearrangement of $\mathtt{SWAP}$ gates. (b)~$\mathtt{SWAP}$-gate equivalence rules mapped onto spin configurations. Notice the upper and lower spin-down rows to enforce the idle-qubit condition.}
    \label{fig:rules}
\end{figure}

\paragraph{Compilation as a thermodynamic process.}We represent circuits as configurations of a many-body system~\cite{Rattacaso2025QQCompiling}. Since circuits realizing the same unitary can be grouped into equivalence classes $\mathcal{C}_\mathcal{U}$, the space of circuits is the disjoint union of these equivalence classes, i.e., $\mathcal{C}~:=~\bigcup_{\mathcal{U}}\mathcal{C}_\mathcal{U}$. The infidelity cost function defines an energy for this many-body system, and compiled circuits correspond to the ground states in each equivalence class.

Given an initial uncompiled circuit, we define the thermal equilibrium state for the system as the probability distribution of equivalent circuits that minimizes the free energy $F(A, T) = I(A) - TS$, where $S$ is the entropy, $T$ is the temperature, and $I$ is the infidelity cost function defined in Eq.~\eqref{eq:infidelity}. In this framework, quantum-circuit compilation is a thermodynamic process in which one starts from a high-temperature state and converges to an equal-probability distribution over optimal equivalent circuits by slowly reducing the system temperature.

We simulate this thermodynamic process by introducing an Equivalence-Class Simulated Annealing (ECSA), i.e., a temperature-driven Markov chain Monte Carlo (MCMC) algorithm restricted to the states representing equivalent circuits.  This is obtained by using a set of local equivalence rules $\mathcal{E}$ as an update rule in the MCMC. The rules in $\mathcal{E}$ depend on the chosen gate set $\mathbb{G}$ and form a sound and complete equational theory~\cite{Clement2023EqReasQC, Clement2024EqReasQCMinimal}: all and only equivalent quantum circuits can be derived by applying the rules to the input circuit. Typical examples of equivalence rules include exchanging commuting gates, shifting gates into previously idle layers, and removing consecutive self-inverse gates (see first rule in Fig.~\ref{fig:rules}a). As a result of this choice of the update rules, the states sampled by the ECSA always encode equivalent quantum circuits~\cite{Rattacaso2025QQCompiling} (see Sec.~\ref{app:sim_anneal} of the Supplemental Material for details about the implementation of the ECSA).

We link the emergence of phase transitions with structural features of the compiled quantum circuits. In particular, we find that during the SA, the system crosses Ising-like phase transitions, and that the optimal circuit belongs to spin-density-wave phases that depend on the Hamiltonian parameters. 
It is crucial to note that these phases emerge only by restricting the MCMC to the equivalence class of the initial circuit. 
Without this constraint, for $i_{\mathtt{G}} > i_{\mathtt{Idle}}$, the trivial empty circuit minimizes the infidelity cost function in Eq.~\eqref{eq:infidelity} independently of the initial circuit. 

\paragraph{Quantum circuits as 2D spin lattices.} The complexity of the compilation process is primarily driven by two-qubit gates; in their absence, the problem factorizes into independent single-qubit subproblems. For this reason, we focus first on circuits containing only two-qubit gates, while afterward we generalize our findings. Within this hypothesis, we define a lattice in which each site corresponds to a link between neighboring qubits at time $t$. We also assume a one-dimensional geometry with nearest-neighbor connectivity for the qubit register, but it can be generalized to higher dimensions.

We rewrite the infidelity Hamiltonian as a 1/2-spin Hamiltonian. To this purpose, a circuit $\mathbf{C}$ containing only two-qubit gates can be mapped onto a spin configuration $\boldsymbol{\sigma}(\mathbf{C}) = \{\sigma_{(t,b)}\}$ on a lattice of size $N_t \times (N_q - 1)$. We assign a spin variable $\sigma_{(t,b)} = \pm 1$ to each link $b$ between neighboring qubits in the circuit, such that $\sigma_{(t,b)} = +1$ indicates a gate applied to qubits $b$ and \mbox{$b+1$}. Notice that, according to this mapping, a qubit $q$ is $\mathtt{Idle}$ if and only if both spins $b = q - 1$ and $b = q$ are down. As a consequence, when mapping equivalence rules, we add spin-down sites where needed to enforce the $\mathtt{Idle}$ qubit condition (see the upper and lower spin-down rows in Fig.~\ref{fig:rules}b). As shown in Sec.~\ref{app:hamiltonian} of the Supplemental Material, the Hamiltonian cost in Eq.~\eqref{eq:infidelity} can be rewritten as
\begin{widetext}
    \begin{equation}
    \label{eq:hamiltonian}
    H_{\mathrm{infidelity}} = \sum_{t}\left[\left(i_{\mathtt{G}} - 2i_{\mathtt{Idle}}\right)\sum_{b=1}^{N_q-1}n_{(t,b)} + \sum_{b=1}^{N_q-1}\sum_{b'=b+2}^{N_q-1}\left(\dfrac{A}{b' - b}\right)^6 n_{(t,b)}n_{(t,b')}-N_qi_{\mathtt{Idle}}\prod_{b=1}^{N_q-1}\left(1-n_{(t,b)}\right)\right],
\end{equation}
\end{widetext}
where $n_{(t,b)} = \frac{\sigma_{(t,b)}+1}{2}$ is the two-qubit gate density operator.
The local density term accounts for the cost of a two-qubit gate $i_\mathtt{G}$, removing twice the cost of an idle qubit. The crosstalk infidelity is represented as a long-range, density-density repulsive term.
Notice that this interaction is not isotropic in the effective lattice, as it acts only between sites at the same time layer $t$. Finally, the third term removes the cost of having fully empty layers, which should not be taken into consideration when computing the overall infidelity as they can be skipped. While the Hamiltonian includes explicit interactions only along the qubit-link direction, the circuit-equivalence rules induce effective inter-layer couplings. Consequently, the equivalence-constrained ensemble behaves as a 2D interacting spin system.

\begin{figure*}
    \centering
    \includegraphics[width=\linewidth]{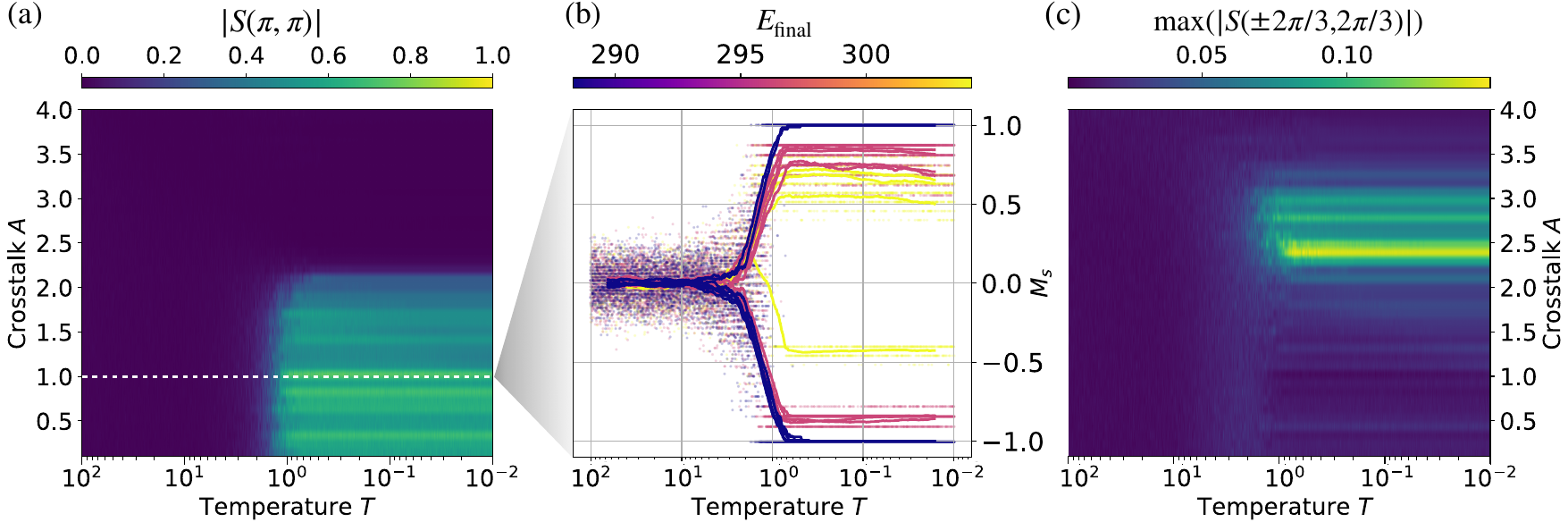}
    \caption{$\mathbb{Z}_2$ symmetry-breaking phase during compilation of a reversal permutation circuit. (a)~Structure factor at $\vec{k} = (\pi, \pi)$ for a circuit with $N_q = 8$ qubits, averaged over 20 Monte Carlo samples, as a function of the temperature $T$ and the crosstalk radius $A$. (b)~Staggered magnetization of the corresponding link-spin lattice for the 8-qubit circuit at fixed $A = 1$ (dashed line on Fig.~\ref{fig:fullperm_z2}a), plotted for all Monte Carlo samples and smoothed with a moving average. The color assigned to each curve represents the final circuit energy at $T \approx 0$. (c)~Absolute value of the structure factor at $\vec{k}=\left(\pm\frac{2\pi}{3}, \frac{2\pi}{3}\right)$ for a circuit with $N_q = 8$ qubits, averaged over 20 Monte Carlo samples, as a function of the temperature $T$ and the crosstalk radius $A$. Color fluctuations in the ordered region are due to the presence of low-temperature $\mathbb{Z}_3$-breaking configurations sampled during the ECSA process.}
    \label{fig:fullperm_z2}
\end{figure*}

\paragraph{Ising phase transition.} We compute the phase diagram of the Hamiltonian $H_{\mathrm{infidelity}}$ within equivalence-constrained ensembles as a function of temperature and crosstalk radius. 
We fix $i_{\mathtt{Idle}} = 1$ and $i_{\mathtt{G}} = 10$ and generate 20 independent Monte Carlo processes for each value of the crosstalk radius $A$. While lowering the temperature, we sample spin configurations. For each sample, we remove empty time layers, as they are nonphysical, so the effective number of time layers is $ \tilde{N}_t = N_t - N_t^{\mathrm{empty}}$.
To characterize the configuration distributions, we measure the magnetization $M_{t,b} = \langle\sigma_{(t,b)}\rangle$ and the static structure factor $S(\vec{k}) = \frac{1}{\mathcal{V}^2}\sum_{t,\,b}\sum_{t',\,b'} \langle\sigma_{(t,b)}\sigma_{(t',b')}\rangle e^{-i\vec{k}\cdot \vec{r}}$, where $\langle\cdot\rangle$ indicates the
average over all the configurations at fixed $T$ and $A$ and $\mathcal{V}$ is the effective size of the lattice $(N_q-1)\times \tilde{N}_t$.

First, we consider permutation circuits. They are composed of sequential $\mathtt{SWAP}$ gates that reverse the state of two qubits, namely $\mathtt{SWAP}|q_1q_2\rangle = |q_2q_1\rangle$. 
In Fig.~\ref{fig:rules}, we show the set of equivalence rules for $\mathtt{SWAP}$ gates~\cite{coxeter1980generators}. For permutation circuits, the minimum number of required $\mathtt{SWAP}$ gates provides an estimate of the Kolmogorov complexity $\mathcal{K}$, namely the necessary resources to implement a given circuit~\cite{Kolmogorov1968, Kendall1938distance}.
We consider circuits that realize the reversal permutation $[0,1,\dots,N_q-1]\rightarrow[N_q-1,\dots,0,1]$, for which the Kolmogorov complexity is maximal at fixed $N_q$ and amounts to $\mathcal{K}={N_q\left(N_q-1\right)/2}$. Any other permutation of the same register requires fewer $\mathtt{SWAP}$ and thus is less complex than the reversal permutation circuit. We consider registers with $N_q=[6,8,10]$.
Since simultaneous near $\mathtt{SWAP}$ gates are penalized by the crosstalk, we expect that, in the regime of small crosstalk radius, the minimum error configuration should contain a gate on every other link for each time layer. Moreover, equivalence rules suggest that nontrivial circuits contain sequential $\mathtt{SWAP}$ gates at time $t+1$ spatially shifted by one link with respect to those at time $t$.

At low temperatures and crosstalk radius $A\lesssim2$, the equilibrium circuit has a \textit{brick-wall} shape as the one shown in the blue region of Fig.~\ref{fig:figure1}. Such a configuration translates into an antiferromagnetic link-spin order, giving rise to the ordered phase observed in Fig.~\ref{fig:fullperm_z2}a. 
In panel (b), we plot the staggered magnetization \mbox{$M_s =  \overline{\langle(-1)^{t+b} \sigma_{(t,b)}\rangle} \rvert_{t,b}$} at $A=1$, showing that optimal, equivalent circuits exhibit opposite magnetic order. Each line shows a single annealing process for the same initial circuit, with its color indicating the final energy. Lower energies correspond to larger magnetization, confirming that the optimal circuit is an antiferromagnet. The emergence of a $\mathbb{Z}_2$-symmetry breaking within equivalence suggests a second-order phase transition.
To verify this hypothesis, we consider the structure factor $|S(\pi, \pi)|$ as the order parameter to perform a finite-size scaling analysis. By using the ansatz $\lvert S(\pi, \pi)\rvert L^{\frac{2\beta}{\nu}} = f\left((T - T_c)L^{\frac{1}{\nu}}\right)$, with $L = N_q - 1$~\cite{Onsager1944Crystal, Kadanoff1967CritPoints, Wilson1974RG, Nishimori2010PTCriticalPhenomena,Kadanoff1966ScalingIsing}, we find $T_c \approx 1.01 \pm 0.05$, $\beta \approx 0.09 \pm 0.03$, $\nu \approx 1.14 \pm 0.16$, which are consistent with the 2D Ising universality class (for details, see Sec.~\ref{app:fss} of the Supplemental Material).

We now investigate regimes at higher crosstalk radius $2 \lesssim A \lesssim 3$.
In analogy with the antiferromagnetic case described above, we observe a $\mathbb{Z}_3$-ordered regime in which the gates are separated by two links (see the circuit in the orange region of Fig.~\ref{fig:figure1}). 
To define the boundaries of this regime, we take as the order parameter the structure factor in $2\pi/3$, as shown in panel (c) of Fig.~\ref{fig:fullperm_z2}. While the sizes of the circuits we address here are not sufficiently large to verify higher-order $\mathbb{Z}_n$ phases, we expect such orders to emerge at higher crosstalk radius. For finite register sizes, we expect the circuit to become fully serial at a sufficiently large radius $A \gg N_q$, namely, each layer contains at most one gate.

\paragraph{Random permutations.} 
We consider now random-permutation circuits composed of $\mathtt{SWAP}$.  
At lower Kolmogorov complexity, we find that the $\mathbb{Z}_2$ order is still visible in all cases, with the maximal value of the structure factor in the ordered phase increasing with the Kolmogorov complexity (see Sec.~\ref{app:randomperm} of the Supplemental Material). This marks a tight connection between emergent order and the complexity of the compiled circuit, as more complex permutation circuits yield stronger collective ordering under the equivalence constraint.

\begin{figure}
    \centering
    \includegraphics[width=\linewidth]{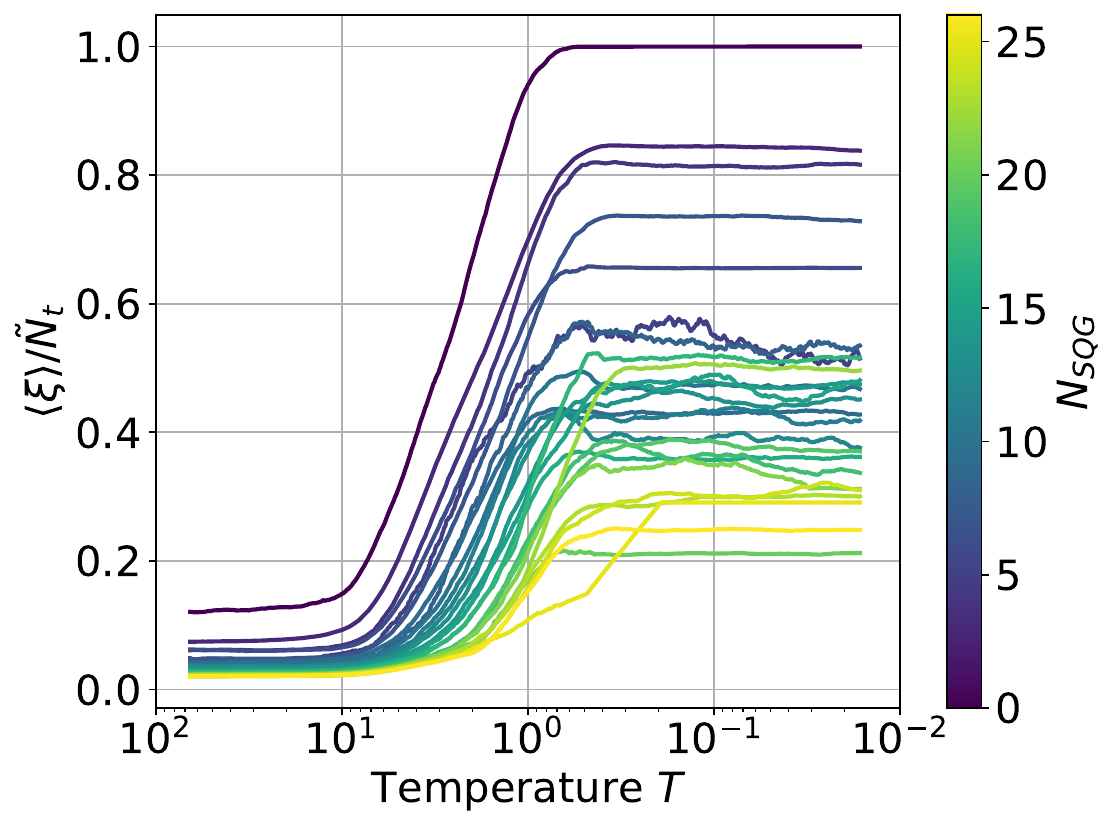}
    \caption{Normalized mean size of antiferromagnetic clusters along the layer direction $\langle\xi\rangle/\tilde{N}_t$ for 6-qubit reversal permutation circuits doped with single-qubit gates randomly selected from $\{\mathtt{H}, \mathtt{T}\}$, as a function of the annealing temperature. Each curve is obtained by first averaging over Monte Carlo samples converging to the minimum-energy state for each circuit, then averaging over circuits sharing the same number of single-qubit gates after compilation (color bar), and finally applying a moving-average smoothing over 100 points.}
    \label{fig:sqg}
\end{figure}

\paragraph{Generalizing the gate set.} Finally, we consider the effect of adding single-qubit gates to the circuit. Starting from the 6-qubit reversal permutation circuit, we randomly add $N_{SQG}$ single-qubit gates drawn from the set $\{\mathtt{H}, \mathtt{T}\}$. For each initial single-qubit gate count \mbox{$N_{SQG} = 3, 6, 9, \dots, 36$}, we generate and compile 10 random circuit instances, setting the crosstalk radius to \mbox{$A = 1$}. 
The single-qubit gates act as defects in the link lattice, because they force the neighboring links to be down. Therefore, during compilation, we track the mean cluster size rather than the structure factor. In particular, we measure $\langle\xi\rangle/\tilde{N}_t$, where \mbox{$\langle\xi\rangle = \overline{\langle\min\{r: C^s_b(r) < 0\}\rangle}\rvert_{b}$}, with \mbox{$C^s_b(r)=\overline{(-1)^r\sigma_{(t,b)}\sigma_{(t+r,b)}}\rvert_{t}$} being the staggered magnetization autocorrelation function along the time direction.

In Fig.~\ref{fig:sqg}, we plot $\langle\xi\rangle/\tilde{N}_t$, where the line color corresponds to the number of single-qubit gates at the end of compilation. The choice for such an observable is further justified in Sec.~\ref{app:czswap_sqg} of the Supplemental Material. The case  $N_{SQG}=0$ corresponds to the reversal permutation circuit, for which we find $\langle\xi\rangle/\tilde{N}_t = 1$ as expected. As we increase the number of single-qubit gates, the mean cluster size decreases accordingly. This can be understood as single-qubit gates occupying layers that would otherwise be idle, thereby modifying the space of equivalent configurations and increasing the freedom in placing two-qubit gates. From the physical point of view, single-qubit gate defects separate the antiferromagnetic order into multiple clusters whose individual size decreases with defect density. For further details on the scaling of cluster statistics with respect to the number of single-qubit gates and for simulations of random circuits with gate set $\{\mathtt{SWAP}, \mathtt{CZ}\}$, see Sec.~\ref{app:czswap_sqg} of the Supplemental Material.

\paragraph{Conclusions and outlook.}In this work, we have linked quantum-circuit compilation with a thermodynamic process, highlighting critical phenomena emerging in a corresponding many-body system. Our findings reveal a tight connection between the emergent phase diagram, computational complexity, and structural features in quantum-circuit compilation. The equivalence constraint among circuits emerges as the fundamental source of complexity in the optimization landscape.

In the future, following the framework introduced in \cite{Rattacaso2026QAlgos4EqReas}, our statistical-mechanical description could be promoted to a quantum one. This would allow us to compare quantum-inspired methods, such as tensor networks and quantum optimization protocols for circuit compilation~\cite{Rattacaso2025QQCompiling}, with classical heuristics through their corresponding phase diagrams, and to investigate whether quantum dynamics can provide a speedup in the optimization procedure. A further extension involves examining how equivalence constraints influence the computational complexity and phase diagram of other optimization problems, such as the Minimum Linear Arrangement Problem~\cite{GONZAGADEOLIVEIRA2026100962}, in which the search space is limited to isomorphic graphs connected by permutations. This issue may be addressed by reformulating the optimization process as a thermodynamic system and analyzing its emergent phases using the ECSA.

\let\oldaddcontentsline\addcontentsline
\renewcommand{\addcontentsline}[3]{}
\begin{acknowledgments}
\paragraph{Acknowledgments.}We thank Rosario Fazio for feedback on the manuscript. The research leading to these results has received funding from the following organizations: the European Union, via project EuRyQa (Horizon 2020, Grant Agreement No.~101070144), project PASQuanS2 (Quantum Technologies Flagship, Grant Agreement No.~101113690), ICSC - Italian Research Center on HPC, Big Data and Quantum Computing (NextGenerationEU Project No.~CN00000013); the Italian Ministry of University and Research (MUR) via: Quantum Frontiers (the Departments of Excellence 2023-2027); the World Class Research Infrastructure - Quantum Computing and Simulation Center (QCSC) of Padova University; Istituto Nazionale di Fisica Nucleare (INFN): iniziativa specifica IS-QUANTUM.
S.N. acknowledges that this project has received funding from the European Union’s Horizon Europe research and innovation program under the Marie Skłodowska-Curie grant agreement No.~101059826 (ETNA4Ryd). We acknowledge computational resources from Cloud Veneto, as well as computational time on Cineca’s Leonardo machine via the project IsB31-QUINOA.
\end{acknowledgments}

\paragraph{Data availability.}The data that support the findings of this article are available upon reasonable request, and have been generated via the software Vulqano~\cite{vulqano}.

\bibliography{bibliography}
\let\addcontentsline\oldaddcontentsline

\clearpage
\onecolumngrid
\begin{center}
    \textbf{\Large Supplemental Material for ``\Title''}
\end{center}

\setcounter{equation}{0}
\setcounter{figure}{0}
\setcounter{table}{0}
\makeatletter
\renewcommand{\theequation}{S\arabic{equation}}
\renewcommand{\thefigure}{S\arabic{figure}}
\setlength\tabcolsep{10pt}
\setcounter{secnumdepth}{2}

\newcommand\numberthis{\addtocounter{equation}{1}\tag{\theequation}}
\newcommand{\insertimage}[1]{\includegraphics[valign=c,width=0.04\columnwidth]{#1}}

\tableofcontents

\section{Equivalence-Class Simulated Annealing}
\label{app:sim_anneal}
In this Appendix, we describe the numerical method used to minimize the infidelity Hamiltonian within the space of spin configurations constrained by circuit equivalence. 

To find low-energy configurations of the infidelity Hamiltonian, we can sample spin configurations from a statistical ensemble that minimizes the free energy. Sampling from the Boltzmann distribution
\begin{equation}
    p \propto e^{-\frac{H}{T}}
\end{equation}
is known to minimize the free energy at temperature $T$. A large $T$ allows a broader exploration of the energy landscape, while as $T \rightarrow 0$ the search becomes more constrained and transitions between local minima become strongly suppressed. 

We implement this Boltzmann sampling through a MCMC SA algorithm, where the update rules correspond to equivalence rules based on the circuit's gate set. At each step $k$ of the Markov chain, an update rule is proposed and accepted with a probability controlled by the temperature $T_k$ at that step. Given an initial configuration $\boldsymbol{\sigma}^{\mathrm{in}}(\mathbf{C}^{\mathrm{in}}) \equiv \boldsymbol{\sigma}^{\mathrm{in}}$, associated with an input quantum circuit $\mathbf{C}^{\mathrm{in}}$, and a set of rules $\mathcal{E}$, the algorithm proceeds as follows:
\begin{enumerate}
    \item Randomly choose a subset of spins $\boldsymbol{\sigma}^{\mathrm{in}}_{\mathrm{sub}}$ on which an update rule $\left\{\boldsymbol{\sigma}^{\mathrm{in}}_{\mathrm{sub}} \leftrightarrow \boldsymbol{\sigma}^{\mathrm{out}}_{\mathrm{sub}}\right\} \in \mathcal{E}$ can be applied.
    \item Using the infidelity Hamiltonian, compute the energy of the updated configuration $\boldsymbol{\sigma}^{\mathrm{out}}$ and compare it to the energy of $\boldsymbol{\sigma}^{\mathrm{in}}$.
    \begin{enumerate}[label*=\arabic*.]
        \item If the energy decreases, we accept the change, apply the rule, and move to the next step, using the updated configuration as the new input.
        \item If the energy increases, we accept the change with probability $p(k) = e^{-\frac{\Delta E}{T_k}}$, with $\Delta E$ the energy difference. If rejected, the input configuration remains unchanged.
    \end{enumerate}
\end{enumerate}
The temperature at each step follows an annealing schedule. In this work, we use
\begin{equation}
\label{eq:schedule}
    T_k = T_{\mathrm{max}}\left(\frac{T_{\mathrm{min}}}{T_{\mathrm{max}}}\right)^{k/N_{\mathrm{steps}}}
\end{equation}
with $k = 0, \dots, N_{\mathrm{steps}}$. A larger number of annealing steps $N_{\mathrm{steps}}$ slows down the cooling rate, improving convergence to the true ground state. In all simulations, we fix the temperature bounds to $T_{\mathrm{max}} = 100$ and $T_{\mathrm{min}} = 0.01$, while the number of steps is varied according to the size of the simulated circuit. The initial and final temperatures are chosen based on the maximum and minimum energy jumps allowed by the rule, so that at the beginning of the annealing process, any move is accepted with high probability, while at the end, only energy-decreasing moves are accepted.

The choice of the input circuit $\mathbf{C}^{\mathrm{in}}$ is arbitrary within the equivalence class associated with the unitary implemented by the circuit. To avoid biased sampling, the state at the beginning of the process must be sampled from the equal-probability distribution that encodes the infinite-temperature ensemble of equivalent circuits. To this aim, starting from any circuit in the equivalence class, we generate the input circuit $\mathbf{C}^{\mathrm{in}}$ by running the MCMC algorithm at a very high, fixed temperature and a free Hamiltonian $H = 0$, allowing a free exploration of the equivalence class. Finally, we run the MCMC algorithm with the annealing schedule from Eq.~\eqref{eq:schedule} and with the input circuit $\mathbf{C}^{\mathrm{in}}$ chosen as a random circuit within the desired equivalence class.

\section{Derivation of the infidelity Hamiltonian in the link spin mapping}
\label{app:hamiltonian}
In this Appendix, we detail the derivation of the Hamiltonian in the link spin mapping (Eq.~\eqref{eq:hamiltonian}) starting from the infidelity cost function $I(\mathbf{C})$ (Eq.~\eqref{eq:infidelity}).

We divide our derivation in three different steps, each corresponding to a different contribution in the overall infidelity. In the following, we define $n_{(t,b)} := \dfrac{\sigma_{(t,b)}+1}{2}$.
\begin{enumerate}
    \item \textbf{Infidelity of a gate}. Assuming for simplicity that all gates yield the same infidelity $i_\mathtt{G}$, it is sufficient to count the number of spin-up sites in the circuit. The number of gates $N_\mathtt{G}$ can be easily estimated as the sum of the number operator $n_{(t,b)}$ for all the lattice sites. Then, the Hamiltonian takes the form
    \begin{equation}
        H_{\mathtt{G}} = i_{\mathtt{G}} \sum_t \sum_b n_{(t,b)}.
    \end{equation}
    \item \textbf{Infidelity of an idle qubit}. While a spin-up always corresponds to the presence of a two-qubit gate, the number of spin-down sites cannot be directly used to count the number of idle qubits. Nevertheless, this number can be estimated from the number of two-qubit gates as
    \begin{equation}
        N_{\mathtt{idle}} = N_q - 2N_{\mathtt{G}}.
    \end{equation}
    Consequently, the infidelity Hamiltonian for idle qubits reads
    \begin{align}
        H_{\mathtt{Idle}} &= i_{\mathtt{Idle}}\sum_t (N_q - 2\sum_bn_{(t,b)}) \\ &= N_q i_{\mathtt{Idle}} N_t - 2i_{\mathtt{Idle}}\sum_t\sum_bn_{(t,b)}. \nonumber
    \end{align}
    Since the lattice dimensions are fixed, idle layers may form during compilation. These layers are not meant to be executed during circuit operation. Despite that, they do contribute to the infidelity cost in the expression above. Hence, it is necessary to introduce an additional energy term to remove their contribution. The number of empty (idle) layers $N_t^{\mathtt{empty}}$ can be estimated as
    \begin{equation}
       \label{eq:n_t_empty}
       N_t^{\mathtt{empty}} = \sum_t \prod_b (1-n_{(t,b)}).
    \end{equation}
    By subtracting the contribution $N_q i_{\mathtt{Idle}} N_t^{\mathtt{empty}}$ of empty layers, the final form of the infidelity Hamiltonian for idle qubits becomes
    \begin{equation}
        H_{\mathtt{Idle}} = N_qN_ti_{\mathtt{Idle}}-N_q i_{\mathtt{Idle}}\sum_t\prod_b(1 - n_{(t,b)}) - 2i_{\mathtt{Idle}}\sum_t\sum_bn_{(t,b)}\nonumber\,.
    \end{equation}

    \item \textbf{Crosstalk}. Since configurations where two spin-up sites are placed in the same layer $t$ at neighboring links $b$, $b+1$ do not correspond to any quantum circuit, we consider crosstalk between spin-up sites at a distance $\lvert b'-b \rvert \geq 2$. From the infidelity cost function, we can directly construct a crosstalk infidelity Hamiltonian as
    \begin{equation}
        H_{\mathrm{crosstalk}} = \sum_t \sum_{b=1}^{N_q-1} \sum_{b'=b+2}^{N_q-1}x_{\mathtt{G}\mathtt{G'}}(\lvert b' - b \rvert)n_{(t,b)}n_{(t,b')},
    \end{equation}
\end{enumerate}
with $x_{\mathtt{G}\mathtt{G'}}(\lvert b' - b \rvert) = \left(\dfrac{A}{\lvert b' - b \rvert}\right)^6$.

Summing together all the individual infidelity contributions, the total infidelity Hamiltonian in the link spin mapping reads
\begin{align}
    H_{\mathrm{infidelity}} &= H_{\mathtt{G}} + H_{\mathtt{Idle}} + H_{\mathrm{crosstalk}} \nonumber \\
    &= N_qi_{\mathtt{Idle}}N_t + \sum_t {\Bigg[}(i_{\mathtt{G}}-2i_{\mathtt{Idle}})\sum_b n_{(t,b)} +\sum_{b=1}^{N_q-1}\sum_{b'=b+2}^{N_q-1}\left(\dfrac{A}{\lvert b' - b \rvert}\right)^6n_{(t,b)}n_{(t,b')} - N_qi_{\mathtt{Idle}}\prod_{b}(1-n_{(t,b)}){\Bigg]}
\end{align}
Neglecting the first constant energy term, we obtain Eq.~\eqref{eq:hamiltonian}.

\section{Finite-size scaling analysis of the Ising phase transition in quantum-circuit compilation}
\label{app:fss}

In this Appendix, we provide details on the finite-size scaling analysis for the phase transition between uncompiled and maximally parallel circuits at fixed crosstalk radius $A=1$. We consider reversal permutation circuits with increasing qubit register sizes $N_q = [6, 8, 10]$, whose fully serial realizations require $N_t = [15, 28, 45]$ layers, respectively. As evidenced in Fig.~\ref{fig:fullperm_z2}b, optimal circuits correspond to perfectly antiferromagnetic lattices. Therefore, we restrict the analysis to Monte Carlo samples converging to the minimum-energy state, and compute the structure factor by averaging only over this subset of samples. Finally, to remove additional noise from the Monte Carlo sampling, we smoothen the resulting curves with a moving average over a window of 100 points.

\begin{figure}[h]
    \centering
    \includegraphics[width=0.5\linewidth]{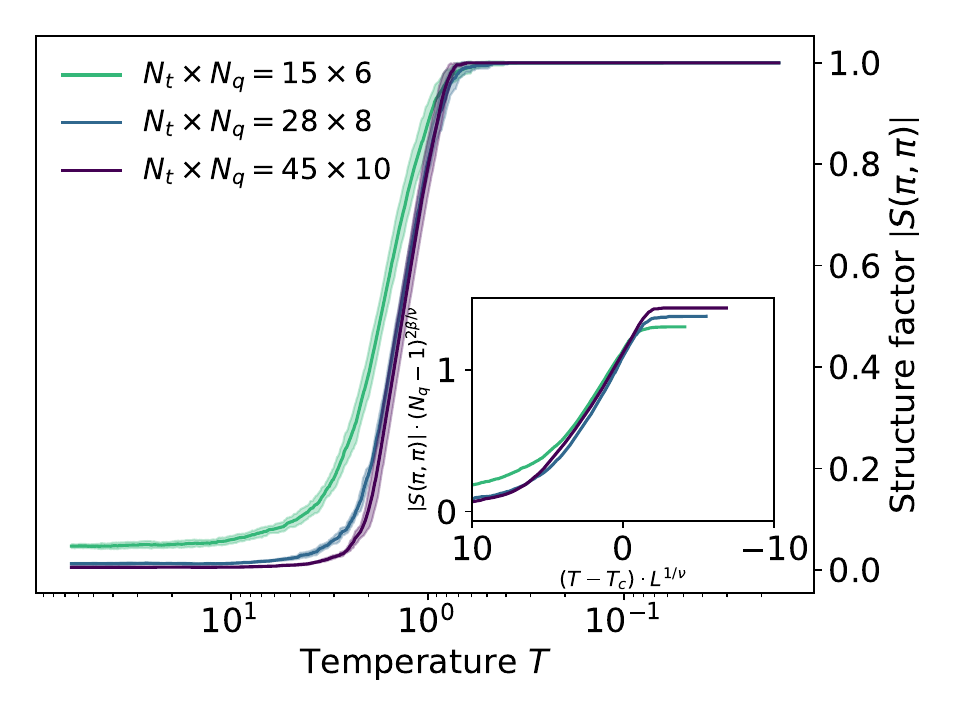}
    \caption{Order parameter $\lvert S(\pi, \pi)\rvert$ at fixed $A = 1$ for increasing circuit sizes, plotted against the annealing temperature. Each curve is obtained by averaging over Monte Carlo samples that converge to the minimum-energy state and applying a moving-average smoothing over 100 points. Inset: universal collapse of the order parameter.}
    \label{fig:fss}
\end{figure}

We attempt finite-size scaling with the ansatz $\lvert S(\pi, \pi)\rvert L^{\frac{2\beta}{\nu}} = f\left((T - T_c)L^{\frac{1}{\nu}}\right)$, where $L = \min(N_q-1, N_t)$. In our case, $L = N_q - 1$, since the spin-lattice representation of the order parameter is defined on the qubit-link degrees of freedom and, for the reversal permutation circuit, $N_q - 1 < N_t$ for all system sizes considered. We obtain the critical temperature $T_c$ and the exponents $\beta$ and $ \nu$ by minimizing a data-collapse cost function that encodes the mean distance among the rescaled structure-factor curves across different system sizes. In particular, for each system size $L$, we fit the rescaled curves $S(\pi, \pi)\rvert L^{\frac{2\beta}{\nu}}$ within a bounded interval around the origin $x \in [-b, b]$, with $x = (T - T_c)L^{\frac{1}{\nu}}$, using a polynomial interpolation $p_L(x; T_c, \beta, \nu)$ up to the fourth degree. The result of the fit directly depends on the choice of the critical parameters $T_c, \beta, \nu$. We then evaluate the fitted curves on a uniform grid of $n_x = 100$ points $\left\{x_j\right\}_{j=1}^{n_x} \in [-b, b]$, with $b = 5$. The optimal parameters can then be written as
\begin{equation}
    (\hat{T}_c, \hat{\beta}, \hat{\nu}) = \argmin_{T_c, \beta, \nu} \sum_L\sum_{L' > L} \left[\frac{1}{n_x}\sum_{j=1}^{n_x} \left(p_{L}(x_j; T_c, \beta, \nu) - p_{L'}(x_j; T_c, \beta, \nu)\right)^2\right]^{1/2}
\end{equation}
We repeat this optimization for a large range of initial guesses for each critical parameter, and post-select the 100 best solutions based on the cost function. Fig.~\ref{fig:fss} shows the curves of the order parameter for increasing system sizes as a function of the annealing temperature. The inset shows the rescaled curves with the critical parameters extracted from our optimization procedure, namely $T_c \approx 1.01 \pm 0.05, \beta \approx 0.09 \pm 0.03, \nu \approx 1.14 \pm 0.16$.

\section{Random permutations}
\label{app:randomperm}

\begin{figure}[ht]
    \centering
    \includegraphics[width=\linewidth]{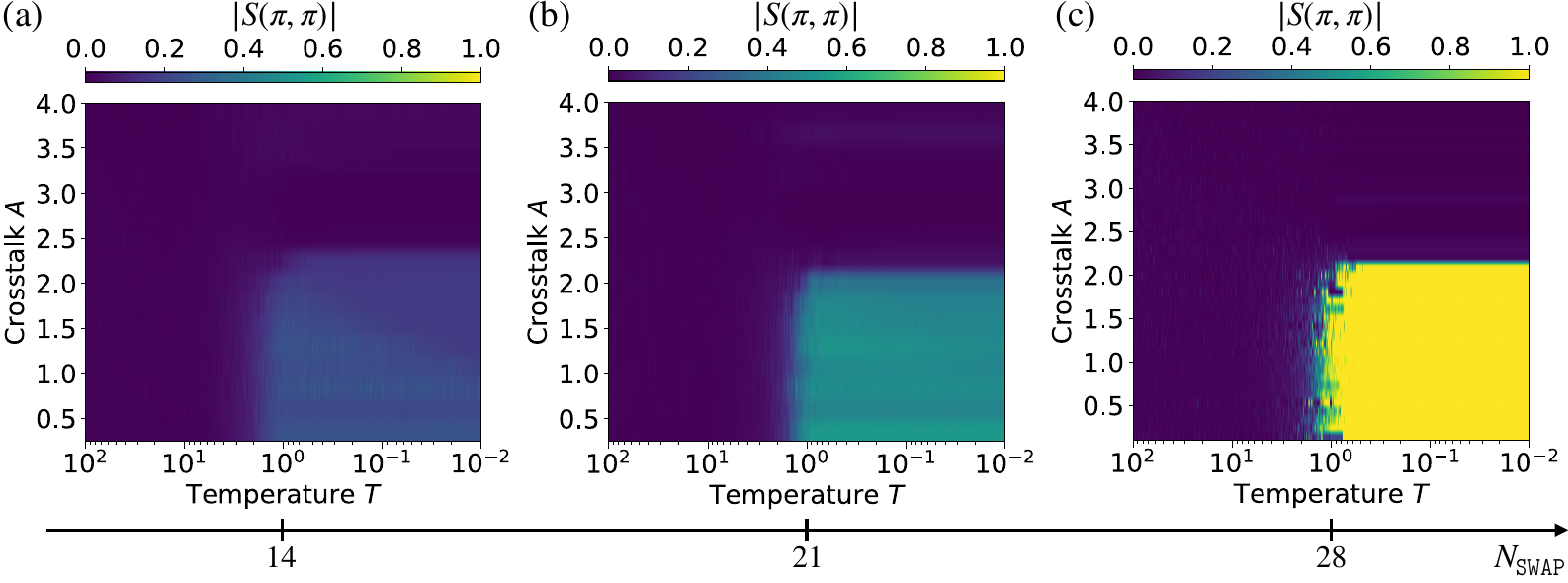}
    \caption{Structure factor at $\vec{k} = (\pi, \pi)$ as a function of the temperature $T$ and the crosstalk radius $A$ for circuits with $N_q = 8$ qubits implementing permutations with Kolmogorov complexities (a)~14, (b)~21, and (c)~28. The order parameter is averaged over Monte Carlo samples that converge to the minimum-energy state.}
    \label{fig:randomperm}
\end{figure}

In this Appendix, we provide evidence for the emergence of $\mathbb{Z}_2$ order in circuits implementing random permutations. The complexity of a permutation $\{\pi(i)\}_{i=0}^{N_q - 1}$ is quantified by its Kendall tau distance from the original qubit ordering, defined as the number of inverted pairs:
\begin{equation}
    \mathcal{K}(\pi) = \left|\left\{(i,j):\, i<j,\ \pi(i)>\pi(j) \right\}\right|.
\end{equation}
For a fixed register size $N_q$, the reversal permutation maximizes this distance and therefore represents the most complex permutation circuit in terms of $\mathtt{SWAP}$ gate count. For $N_q = 8$, the reversal permutation requires a minimum of 28 $\mathtt{SWAP}$ gates. We therefore compare the phase diagram obtained for the reversal permutation circuit with those obtained for random permutations of lower complexities, characterized by Kendall tau distances of 14 and 21. In particular, for each of these complexity values, we generate 10 random permutations with the corresponding Kendall tau distance and compile the associated permutation circuits with ECSA. 

Fig.~\ref{fig:randomperm} compares the resulting phase diagrams for increasing Kolmogorov complexities, measured through the structure factor at $(\pi, \pi)$ averaged over Monte Carlo samples that converge to the minimum-energy state. Figs.~\ref{fig:randomperm}a and \ref{fig:randomperm}b show the order parameter averaged over 10 permutation circuits with Kolmogorov complexities 14 and 21, respectively. Fig.~\ref{fig:randomperm}c corresponds to the reversal permutation, as it is the only permutation with maximal Kolmogorov complexity. 
\begin{figure}[ht]
    \centering
    \includegraphics[width=0.5\linewidth]{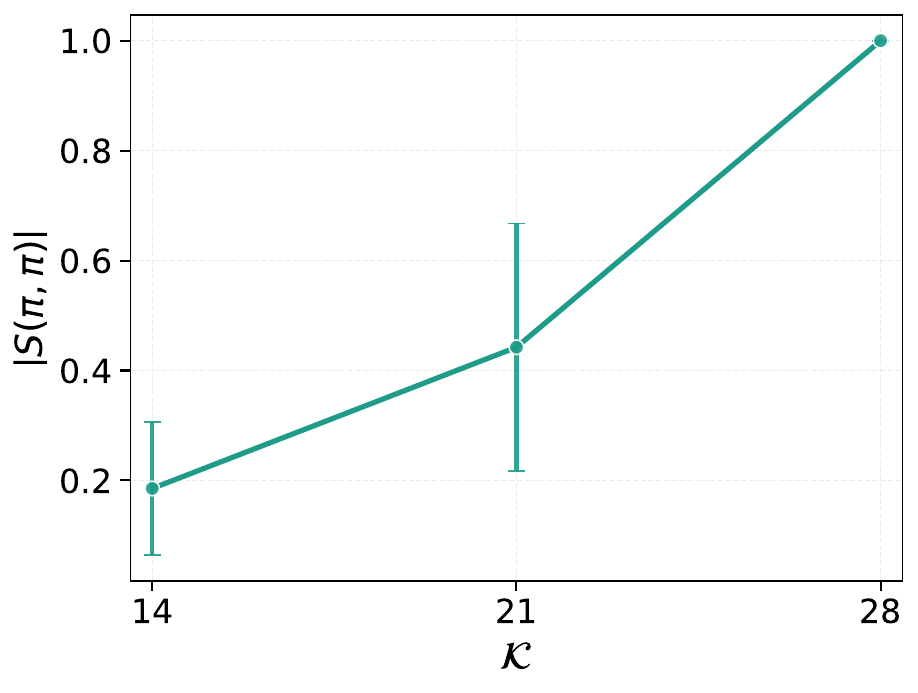}
    \caption{Structure factor $\lvert S(\pi,\pi)\rvert$ at $T \approx 0$ and crosstalk radius $A = 1$ as a function of the Kolmogorov complexity $\mathcal{K}$ for $8$-qubit permutation circuits. For $\mathcal{K} = 14$ and $21$, the points and error bars represent the mean and standard deviation, respectively, over 10 random permutations. The maximal value $\mathcal{K}=28$ is only achieved by the reversal permutation circuit.}
    \label{fig:struc_fac_vs_K}
\end{figure}
The results show that the $\mathbb{Z}_2$-ordered phase remains robust across all considered permutations. Moreover, as shown in Fig.~\ref{fig:struc_fac_vs_K} for $A = 1$, the strength of the ordering, quantified by the peak structure factor, increases with the permutation's Kolmogorov complexity. This trend can be understood from the fact that less complex permutations require fewer $\mathtt{SWAP}$ gates, and, as a result, their maximally parallel realizations generally cannot form a perfect brick-wall circuit. In the spin representation, this manifests as spin-down defects embedded into an otherwise antiferromagnetic lattice, leading to a reduced structure factor.

\section{Further details on circuits with augmented gate sets}
\label{app:czswap_sqg}

In this Appendix, we expand our findings on random circuits with augmented gate sets. We first motivate the order parameter choice for permutation circuits doped with single-qubit gates. We proceed analyzing the effect of the chosen gates $\{\mathtt{H}, \mathtt{T}\}$ on the $\mathbb{Z}_2$ order and formulate a hypothesis on the scaling of the resulting cluster statistics. Finally, we show that random circuits with gate set $\{\mathtt{SWAP}, \mathtt{CZ}\}$ preserve the $\mathbb{Z}_2$-ordered phase.

As stated in the main text, we characterize the effect of single-qubit gates by tracking the mean size of antiferromagnetic clusters in the lattice, as these gates act as defects that divide the antiferromagnetic order. In our model, interactions along the qubit-link direction originate from the crosstalk terms in the infidelity Hamiltonian, whereas interactions along the layer direction are induced by the equivalence rules. Because of this intrinsic anisotropy, the effect of defects along the layer direction is less straightforward than along the qubit-link direction. Moreover, the layer dimension is typically much larger, providing improved statistical sampling. For these reasons, we focus on cluster statistics along the layer direction. 

To quantify the cluster size, we first compute the staggered magnetization autocorrelation function along the layer direction for a fixed qubit link $b$,
\begin{equation}
    C^s_b(r)=\overline{(-1)^r\sigma_{(t,b)}\sigma_{(t+r,b)}}\rvert_{t},
\end{equation}
where the average is taken over all valid layer pairs separated by distance $r$. We then estimate the cluster size as the first distance at which the correlation becomes negative and average this quantity over all qubit links:
\begin{equation}
    \langle\xi\rangle = \overline{\langle\min\{r: C^s_b(r) < 0\}\rangle}\rvert_{b},
\end{equation}
Finally, we normalize this quantity by the number of nonidle layers $\tilde{N}_t$ and track the ratio $\langle\xi\rangle/\tilde{N}_t$ throughout the compilation process.

In our simulations, we consider the single-qubit gate set $\{\mathtt{H}, \mathtt{T}\}$. These two gate species affect the cluster statistics in significantly different ways. Hadamard gates satisfy the identity $\mathtt{H}^2 = \mathtt{Idle}$. Consequently, provided that the two-qubit gate equivalence rules allow them to be placed in parallel with neighboring two-qubit gates, Hadamard gates do not disrupt the antiferromagnetic order of the optimal compiled circuit: they can either be placed in existing layers or transformed in identities pairwise. In contrast, $\mathtt{T}$ gates do not satisfy an analogous cancellation rule. While they satisfy the cyclic relations $\mathtt{T}^2 = \mathtt{S}$, $\mathtt{T}^4 = \mathtt{Z}$, $\mathtt{T}^8 = \mathtt{Idle}$, intermediate products generate distinct single-qubit gate pairs that cannot be eliminated or absorbed into parallel two-qubit gate layers. As a result, consecutive $\mathtt{T}$ gates acting on the same qubit typically introduce additional nonidle layers that divide the antiferromagnetic order into smaller clusters.

\begin{figure}[h]
    \centering
    \includegraphics[width=0.55\linewidth]{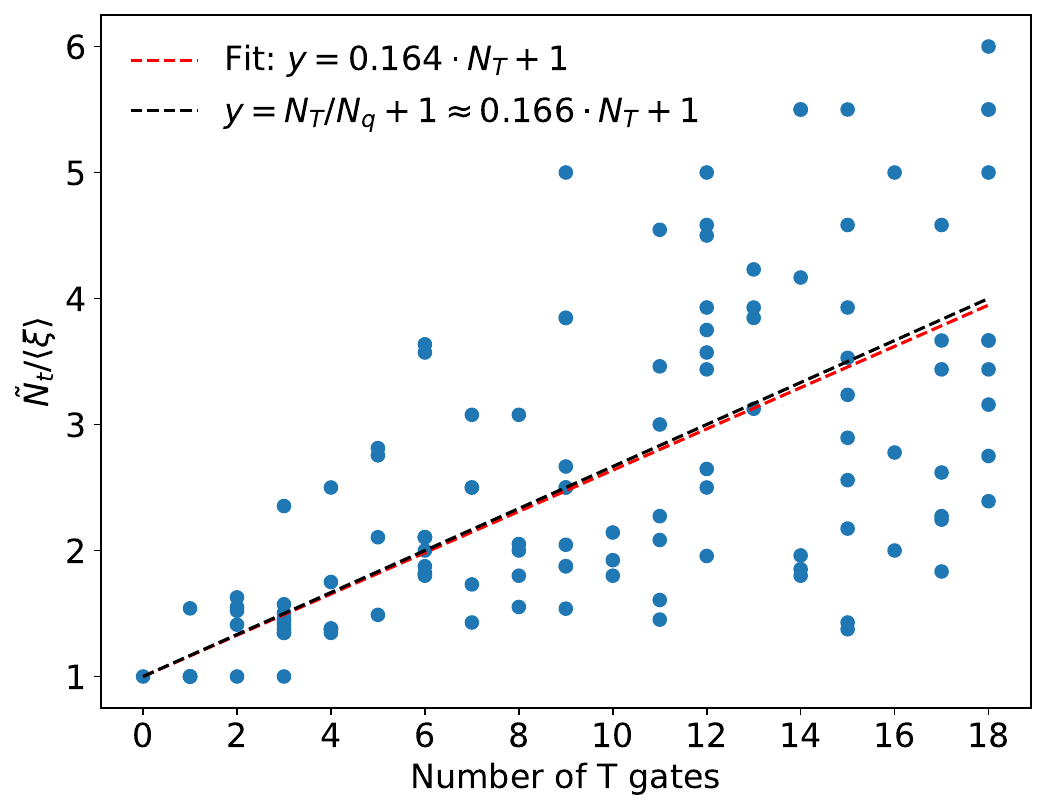}
    \caption{Scatter points: number of antiferromagnetic clusters of compiled 6-qubit reversal permutation circuits doped with single-qubit gates randomly selected from $\{\mathtt{H}, \mathtt{T}\}$, as a function of the number of $\mathtt{T}$ gates in each circuit. Red line: linear fit of the scatter points. Black line: linear function $y = N_\mathtt{T}/N_q + 1$.}
    \label{fig:numclusters_T}
\end{figure}
Based on the previous analysis, we now investigate the scaling of the cluster statistics with the number of single-qubit gate defects. In particular, we estimate the number of antiferromagnetic clusters as $\tilde{N}_t/\langle\xi\rangle$, and study its dependence on the number of $\mathtt{T}$ gates in each compiled circuit instance (Fig.~\ref{fig:numclusters_T}). On average, the number of clusters increases linearly with the number of $\mathtt{T}$ gates, normalized by the qubit register size. Further numerical investigation should be conducted to confirm this scaling hypothesis which is left for future work.

\begin{figure}[h]
    \centering
    \includegraphics[width=0.45\linewidth]{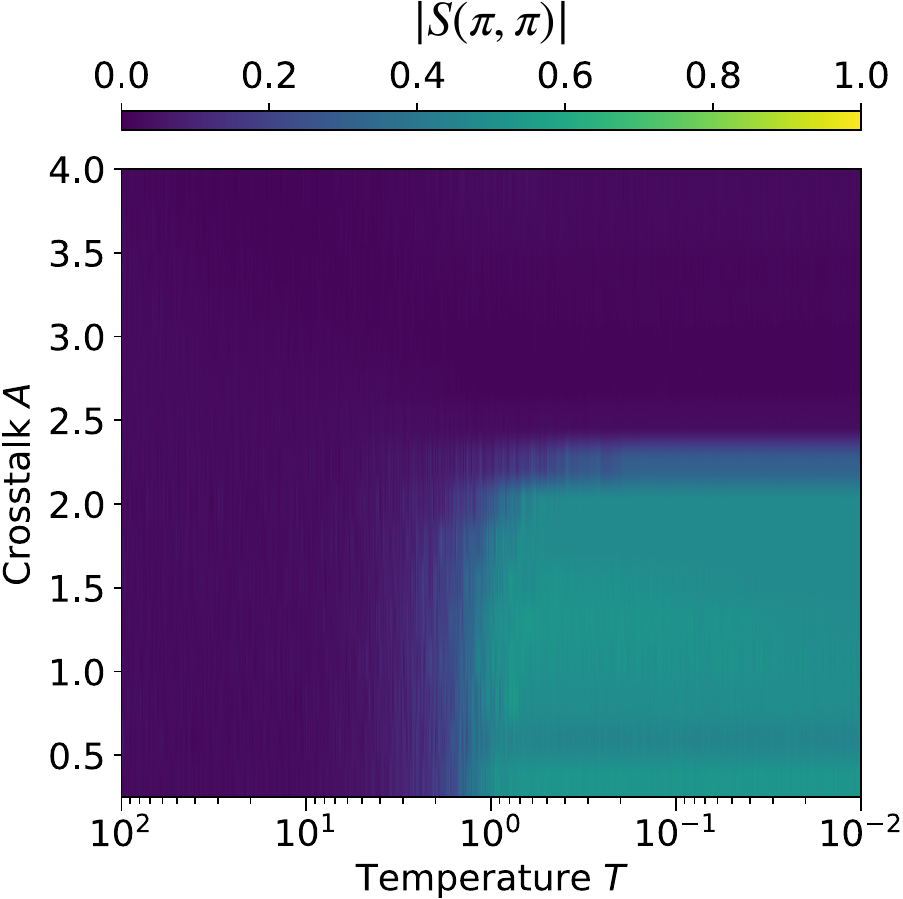}
    \caption{Mean structure factor at $\vec{k} = (\pi, \pi)$ for 10 random circuits with gate set $\{\mathtt{CZ}, \mathtt{SWAP}\}$ with $N_q = 6$ qubits, averaged over 20 Monte Carlo samples, as a function of the temperature $T$ and the crosstalk radius $A$.}
    \label{fig:cz_swap}
\end{figure}

To conclude the analysis of augmented gate sets, we consider random circuits composed only of two-qubit gates from $\{\mathtt{SWAP}, \mathtt{CZ}\}$. We generate 10 random 6-qubit circuits by drawing gates uniformly from this set and placing them in an initial brick-wall structure. While alternative generation procedures might produce circuits with lower Kolmogorov complexity in the worst-case scenario, we have shown in Sec.~\ref{app:randomperm} that the $\mathbb{Z}_2$ ordered phase persists even in lower complexity instances, albeit with a reduced peak of the order parameter. Moreover, the preprocessing step of the ECSA, explained in Sec.~\ref{app:sim_anneal}, removes any bias associated with the initial brick-wall layout before running the actual optimization. 

Compiling these circuits consistently yields configurations with lower energy than the initial brick-wall realizations. In contrast with permutation circuits considered in the main text, the presence of $\mathtt{CZ}$ gates introduces new equivalence rules that allow further gate cancellations during compilation. Fig.~\ref{fig:cz_swap} shows the resulting phase diagram, obtained by averaging the structure factor in $(\pi, \pi)$ over the 10 circuit samples. The persistence of a finite order parameter at low temperature confirms that the $\mathbb{Z}_2$-ordered phase survives in circuits with gate set $\{\mathtt{SWAP}, \mathtt{CZ}\}$.

\end{document}